\documentstyle[manuscript,aps]{revtex}

\newcommand{\beq}{\begin{equation}}
\newcommand{\eeq}{\end{equation}}
\newcommand{\beqa}{\begin{eqnarray}}
\newcommand{\eeqa}{\end{eqnarray}}

\def \veps {\varepsilon}
\def \eps {\epsilon}

\begin{document}
\input epsf

\title{ \bf\Large 
Origin of the Thermal Radiation in a Solid-State\\
 Analog of a
Black-Hole \\ }
\author {  
{\large B. Reznik\footnote{E-mail: reznik@t6-serv.lanl.gov\\
LAUR-97-1055} } \\
{\it Theoretical division, T-6, MS B288, 
Los Alamos National Laboratory}\\
{\it Los Alamos, NM, 87545}\\
{\ } \\
} 
\maketitle 

\begin{abstract}

{An effective black-hole-like horizon occurs,
for electromagnetic waves in matter, at a surface of singular
electric and magnetic permeabilities. In a physical dispersive
medium this horizon disappears for wave numbers with $k>k_c$.
Nevertheless, it is shown that Hawking radiation is still emitted if 
free field modes with $k>k_c$ are in their ground state.} 
\end{abstract}

\vskip 1cm
\vfill\eject
Hawking discovered that black-holes emit 
thermal radiation with temperature $T_H=M_{Planck}^2/8\pi M_{BH}$.
For large black-holes, with mass  $M_{BH}\gg M_{Planck}$,
spacetime curvature is extremely small  near the horizon, and 
the Hawking temperature is much smaller than the Planck
temperature. 
It can be argued therefore,  
that as long as $M_{BH}\gg M_{Planck}$,
the Hawking effect is essentially  a low energy process,
which is decoupled from quantum-gravity effects.
Nevertheless, the  standard  derivation \cite{hawking}
relys on free field theory on curved space-time, and describes 
the  Hawking  radiation as originating from vacuum fluctuations 
on scales exponentially smaller than
$1/M_{Planck}$ \cite{unruh-sonic,jacobson1,jacobson2,englert}.
On such scales, free field theory may not be a valid 
description,
and quantum-gravity may become important.
A question arises, therefore, 
to what extent is the Hawking radiation dependent
on the details of the short distance structure of quantum gravity?
 
In the absence of a theory of quantum-gravity, guiding  
toy models can be helpful. 
Unruh has suggested using a sonic analog of a black-hole 
\cite{unruh-sonic}. In his model, an effective event horizon
is formed for sound waved on a flowing fluid background,
on the surface where the  flow  becomes supersonic.
Unruh has shown, that when  
a natural cutoff is implemented, sonic black-holes
still gives rise to Hawking radiation  \cite{unruh-dumb}.
In his approach the cutoff  is implemented by
modifying  the dispersion relation for sound waves at high frequencies.
This, in turn,
alters the motion of modes with frequency close to the cutoff scale
and gives rise to a new type of trajectories, which approach the
horizon but eventually "reflect" back to infinity. 
It is not clear that a similar process is indeed realized for 
real black-holes.
Further investigations,  based on  Unruh's model 
have been carried out
in  Refs. \cite{bmps,jacobson-origin,corley-jacobson}.

In order to gain further insight into the problem,
in this letter, we  provide and examine,
a different toy model, which is based on 
phenomenological electrodynamics in matter.
It is known that Maxwell's equations
in a medium are formally analogous to 
electromagnetism in curved space-time \cite{schleich}.
In this case, an effective event horizon for 
electromagnetic waves in matter occurs
on a surface of singular electric  and magnetic permeabilities.
As we shall show, formally, thermal radiation can still be emitted,
if the initial state of the field is chosen appropriately.  
Nevertheless, this requires, non-physical, singular electric polarization
on arbitrary short scales, far beyond the scale where
the electromagnetic field and the medium  still couple.
The short distance modifications that must be introduced 
in this toy model are clear:
at high momentum or frequency the medium becomes 
dispersive and gradually the electromagnetic field and the medium 
decouple. This implies that for sufficiently short wave-lengths, 
the effective event horizon disappears.
We shall show however that inclusion of such dispersive behavior
does not eliminate  Hawking radiation. The process which 
generates Hawking's radiation in the present toy model, 
differs however from that found for the sonic model.  
In the following we  adopt units such that 
$\hbar=\kappa_{B}=c=1$.

To begin with, let us consider the 
macroscopic Maxwell's equations in a medium:
\beqa
\nabla \cdot {\bf B}  =  0 \ \ \ \ \ \ & 
\nabla \times {\bf E}+{\partial {\bf B} \over \partial t} =  0  \\
\nabla \cdot {\bf D}  =  0 \ \ \ \ \ \ &
\nabla\times {\bf H} - {\partial {\bf D} \over \partial t} = 0
\eeqa
where ${\bf D}$ and ${\bf H}$ are determined by the
local electric and magnetic polarizations: 
${\bf D} = {\bf E} + {\bf P}$ and ${\bf H} = {\bf B} -
{\bf M}$. 
We shall consider  the special case of linear constitution
relation of the particular form:
\beq
{\bf H}(r,t) = \alpha z {\bf B}(r,t), \ \ \ \ \ \
{\bf D}(r,t) = (\alpha z)^{-1}{\bf E}(r,t), 
\label{constitution}
\eeq
i.e., the electric and magnetic permeabilities are 
$\veps(z)= \mu(z) = 1/(\alpha z)$.

With the gauge choice $\varphi = 0$ and 
$\nabla \cdot \veps{\bf A} =0$, we obtain for the 
vector potential, the equation of motion
\beq
   \alpha^2 z\nabla\times  z \nabla \times {\bf A}
+ {\partial ^2 {\bf A} \over \partial t^2} = 0. 
\label{eq-of-motion}
\eeq
Thus, although space-time is flat, with respect to the rest 
frame of the medium,  waves propagate in an 
effective Rindler geometry:
\beq
ds^2_{effective} =\alpha^2 z^2dt^2 - dz^2 - dx^2 -dy^2.
\eeq
The latter describes the space-time geometry 
near a black-hole horizon, of mass $M_{BH} = (4\alpha)^{-1}$, for 
$r/2M_{BH}\ll 1$ \cite{dbh}.
The event horizon corresponds here to the 
singular surface, $z=0$, where $\veps$ and $\mu \to \infty$.
Although this model clearly breaks-down near the horizon, let us 
first examine its effect on electromagnetic waves
and defer  
the discussion of a more physical realization to the sequel.

The Hamiltonian of the system can be written as
\beq
H = H_I - H_{II} = \int_0^\infty \veps^{-1}(z)
\biggl( {\bf D}^2 + (\nabla\times{\bf A})^2 \biggr)dz 
- \int_{-\infty}^0  |\veps(z)|^{-1}\biggl( {\bf D}^2 + (\nabla\times
{\bf A})^2 \biggr) dz,
\eeq
where ${\bf A}$ and ${\bf D}$ are conjugate canonical 
coordinates.
The energy of a wave is thus negative in the region $z<0$.
This indeed suggests, that like a black-hole, 
the present model can emit thermal radiation.
To see this let us quantize the field.

For simplicity, we shall 
consider in the following 
only waves in the $\hat z$-direction, which 
correspond to radial transverse waves in the black-hole case.
(Since we consider only the case $r/2M_{BH}<<1$, 
the vector nature of electromagnetic waves is unimportant.)  
The solutions of the wave equation (\ref{eq-of-motion}) 
can be written as:
\beq
g_I(\kappa) = \left\lbrace 
\matrix{
{1\over\sqrt{4\pi\omega_\kappa}} e^{-i\omega_\kappa t +i{\kappa
\over \alpha} \ln z}, & 
{\rm for } & z>0 \cr 
0, & {\rm for } & z<0  \cr }
\right\rbrace
\label{g1}
\eeq 
\beq
g_{II}(\kappa) = \left\lbrace 
\matrix{
0, & {\rm for } & z>0  \cr 
{1\over\sqrt{4\pi\omega_\kappa}} e^{i\omega_\kappa t
 +i{\kappa\over \alpha} \ln |z| )}, & {\rm for } & z<0 \cr 
}
\right\rbrace
\label{g2}
\eeq 
where $\omega_\kappa= |\kappa|$. The modes $g_I(\kappa)$ correspond 
to an outgoing (right moving) wave for $\kappa>0$ and ingoing wave for $\kappa<0$.
Similarly, $g_{II}(\kappa)$ corresponds to an outgoing wave for 
positive $\kappa$ and an ingoing wave for negative $\kappa$.
The modes  (\ref{g1},\ref{g2}) 
are orthogonal under the scalar product:
\beq
(g_1,g_2)= i \int_{-\infty}^{+\infty}
\veps(z)  g^*_1 \stackrel{\leftrightarrow}{\partial_t} g_2 dz ,
\label{z-product}
\eeq 
and can be used to express the vector potential as
\beq
A(t,z) = \int_{-\infty}^{+\infty} (g_I(\kappa )a_I(\kappa) +
 g_{II}(\kappa)a_{II}(\kappa) + {\rm h.c.})d\kappa.
\eeq
Here, $a_I$ and $a_{II}$ are the annihilation operators in 
the domains $z>0$ and $z<0$, respectively, 
and $[a_I,a_{II}]=0$.

Next, in order to determine what will be seen by a stationary observer at
$z>0$, we must specify an initial condition for 
the electromagnetic field at the effective horizon. 
Let us define the analogue Kruskal coordinates 
by $U= -ze^{-\alpha t}$ and  $V=ze^{\alpha t}$. In terms of the $U,V$ coordinates
the effective metric is conformally 
flat and the solutions to the wave
equation are  
outgoing right  moving  waves,  $\exp(-i\tilde\omega U)$, 
and ingoing left moving waves, $\exp(-i\tilde\omega V)$.
The Unruh initial condition \cite{unruh76} 
amounts to the requirement
that at the past horizon, $V=0$, the outgoing modes, 
are in their  ground state. 
Thus, time is defined with respect to  
light ray trajectories,  $z = -U e^{\alpha t}$, 
with constant $U$. 
  
The analyticity of the outgoing modes, 
in the lower half of the $U$ complex plane, implies that the combinations 
\beq
f_1(\kappa)=e^{\pi\omega/2\alpha} g_I(\kappa) 
+e^{-\pi\omega/2\alpha}g_{II}^*(-\kappa),
\label{f1}
\eeq
and 
\beq
f_2(\kappa)=e^{\pi\omega/2\alpha} g_{II}(\kappa) 
+e^{-\pi\omega/2\alpha}g_{I}^*(-\kappa),
\label{f2}
\eeq
have only positive frequencies 
with respect to $U$-modes \cite{unruh76}.
It then readily follows that the outgoing state is thermal 
with (global) temperature
\beq
T = {\alpha\over 2\pi}.
\eeq

As in the standard derivation of Hawking radiation, this
model involves ultra-high momenta. To see this, consider an
outgoing photon described by the wave packet
\beq
\int  e^{-(\omega-\omega_0)^2 } e^{-i\omega(t
-{1\over \alpha}\ln z )} d\omega
\propto e^{-(t-{1\over \alpha}\ln z)^2/4}.
\label{packet}
\eeq
As we evolve the wave backward in time, 
the wave packet approaches the horizon like 
$z \approx \exp \alpha t$, and gives rise to ultra-high momentum
$ k \approx \omega \exp(-\alpha t)$. Clearly this phenomena
is also related to the occurrence of singular 
polarization on the horizon.

Nevertheless, since the present model involves known
physics, we do have some handle on these problems. 
When the wave length becomes comparable to 
the ``molecular'' length scale,  we will 
have to account for the 
dispersive properties of the medium, which so far have been 
ignored. 
In particular, for sufficiently large
wave numbers, the electromagnetic waves 
decouples from the medium degrees of freedom,  
and the  effective geometry for the electromagnetic field becomes
in this limit flat. 
Since in our model, the momentum $k$ increases near the horizon, 
while the frequency, $\omega$, remains constant,  
we will introduce {\it spatial-dispersion} 
into the constitution equations (\ref{constitution}).
As we shall see,  this  indeed eliminates the above
mentioned difficulties.

In order to describe both the inhomogeneity 
(in the $z$ direction) and spatial
dispersion, we shall 
now assume the following constitution relation for the electric 
field:
\beq
{ D}(z,t) = \int  \veps(z,z-z') {E}(z',t)dz' .
\label{dispersion-relation}
\eeq 
and a similar relation, with $\mu(z,z-z')=\veps(z,z-z')$, 
for the magnetic field.
The first argument in $\veps(z,z-z')$ allows for spatial
non-homogeneity of the electric polarization, 
which in turn induces a low-momentum effective Rindler geometry. 
The second argument gives rise to a spatial non-locality around 
the point $z$, 
which is  taken to be homogeneous, i.e. to depends only on  $z-z'$.
This allows us to simplify eq. (\ref{dispersion-relation})
by going to the 
momentum representation. Assuming further 
that $\veps$ is analytic in 
its first argument, we obtain:
\beq
{ D}(z,t) = \int  e^{ikz} \veps(i\partial_k, k) 
{ E}(k,t) dk, 
\eeq 
where
\beq
\veps(i\partial_k, k) = \int e^{-i kz} \eps (i\partial_k, z)dz,
\eeq
and ${ E}(k,t)$ is the Fourier transform  of ${ E}(z,t)$.

Using a similar relation between the magnetic field and
the magnetic induction,  we are led to the 
Hamiltonian  
\beq
H = \int  \biggl(  D(k,t)\veps^{-1}(i\partial_k,k){D}(k,t)
+ kA(k,t)\veps^{-1}(i\partial_k,k)k A(k,t)\biggr)dk.
\eeq
Here, $D(k,t)$ and $A(k,t)$ are the canonical coordinates.
The equation of motion for the vector potential 
becomes
\beq
\veps^{-1}(i\partial_k,k) k \veps^{-1}(i\partial_k,k) k 
 A(k) - \omega^2  A(k)\ = 0,
\label{general-eq-of-motion}
\eeq
where $A(k,t) = \exp(\pm i \omega t) A(k)$.
Eq. (\ref{general-eq-of-motion}) can be 
further simplified, by noting that the solutions satisfy
\beq
\veps^{-1}(i\partial_k,k) k A = \pm \omega  A.
\label{firstorder}
\eeq
The latter has a natural interpretation
of an ``eigenvalue-like'' equation for the 
differential dielectric operator $\veps^{-1}$.
 
We shall now make some assumptions on the behavior of $\veps^{-1}$.
For $k<<k_c$, where $1/k_c$ is the ``molecular'' length scale,
we must recover the non-homogeneous dependence
$\veps =1/\alpha z$, thus: $\veps^{-1}(i\partial_k, k<<k_c)\to
i\alpha\partial_k$. On the other hand, for $k>>k_c$, 
we expect the electromagnetic field  
to decouple from the medium, thus:
$\veps^{-1}(i\partial_k,k>>k_c)\to (\epsilon_c)^{-1}$, where 
$\epsilon_c$ is a constant.
Assuming that $\veps^{-1}$ interpolates  smoothly between these two 
limiting cases, we shall represent the dielectric operator by
\beq
\veps^{-1}(i\partial_k, k) = \tilde \theta(k_c-k)
i\alpha\partial_k + \tilde\theta(k-k_c)\epsilon_c^{-1},
\label{dispersion}
\eeq
where $\tilde\theta(k)\equiv {1\over2}[1+\tanh(k/\Delta)]$
is the Heaviside step-function,  and $\Delta$ determines 
the width of the transition.

With this choice the solutions of eq. (\ref{general-eq-of-motion}) 
are given by:
$$
G_\eta(k,t) ={1\over  k}\exp\biggl\{ {i\over\alpha\epsilon_c}
\int_{k_0}^k\Big[ e^{2(k'-k_c)/\Delta} -
{\eta\epsilon_c\over k'}(1 + e^{2(k'-k_c)/\Delta})\Bigr]dk' 
\biggr\}e^{\pm i\omega_\eta t}
$$
\beq
= {1\over k} \exp\biggl\{ {i\Delta\over2\alpha\epsilon_c}
\Bigl[e^{2(k-k_c)/\Delta}-e^{2(k_0-k_c)/\Delta}\Bigr]
-i{\eta\over \alpha}\ln(k/k_0) - i{\eta\over \alpha} e^{-2k_c/\Delta}
\Bigl[{\rm Ei}({2\over\Delta}k) - {\rm Ei}({2\over\Delta}k_0)\Bigr]
\biggr\}e^{\pm i\omega_\eta t},
\label{exact}
\eeq
where ${\rm Ei(x)}=\int_{-\infty}^x (e^t/t)dt$ is the 
integral-exponential function, and $\omega_\eta=|\eta|$.

For  $k<<k_c$ (\ref{exact}) reduces to 
${1\over k}\exp( \pm i\omega_{\eta} t)\exp(-i{\eta\over\alpha}\ln|k|)$.
Substituting this into the Hamiltonian, we find that 
in the limit of $k<<k_c$, 
$\eta k>0$ and $\eta k<0$ correspond to positive and 
negative energy modes, respectively.
In  the other limiting case, $k>>k_c$, 
we have free waves 
with $\omega = |k|/\epsilon_c$, and the energy is  
positive for $\omega>0$.

In order to relate these solutions with the modes 
(\ref{g1},\ref{g2}), and in particular with the 
thermal modes (\ref{f1},\ref{f2}), 
it will be useful to define the following
low momentum modes:
\beq
\tilde g_I(\eta) = \left\lbrace 
\matrix{
{1\over\sqrt{4\pi\omega_\eta}}k^{-1}
e^{-i\omega_\eta t - i \eta {1\over\alpha}\ln k}, & 
{\rm for } & k>0 \cr 0, & {\rm for } & k<0  \cr }
\right\rbrace,
\label{kmode1}
\eeq 
\beq
\tilde g_{II}(\eta) = \left\lbrace 
\matrix{
0, & {\rm for } & k>0  \cr 
{1\over\sqrt{4\pi\omega_\eta}} 
k^{-1} e^{i\omega_\eta t - i\eta  {1\over\alpha}\ln |k|)}, & 
{\rm for } & k<0 \cr 
}
\right\rbrace,
\label{kmode2}
\eeq 
for positive $\eta$. For negative $\eta$ we replace 
above $\omega \to -\omega$.
These modes form in the limit $k_c\to \infty$, 
an orthogonal set under the scalar product:  
\beq
(\tilde g_1,\tilde g_2)= i \int^{+\infty}_{-\infty}
 dk \tilde g^*_1\veps(i\partial_k,k)\stackrel{\leftrightarrow}
{\partial_t}  \tilde g_2 . 
\label{k-product}
\eeq 
For finite $k_c$,  
we need to replace  $\tilde g_{I,II}$  
with the exact solutions $G_\eta(k,t)$.
In the domain $|k|<k_c$, however, we can now expand the field as:
\beq
{A}(t,k) = \int_{-\infty}^{+\infty}
 (\tilde g_I(\eta)\tilde a_I(\eta) + 
\tilde g_{II}(\eta) \tilde a_{II}(\eta) + {\rm h.c.})d \eta, 
\eeq
where $\tilde a_I$ and $\tilde a_{II}$ are the annihilation operators for  
$k>0$ and $k<0$, states respectively. 

Next let us  notice that although the spectrum 
of  $g_I$ or $g_{II}$ in eqs. 
(\ref{g1},\ref{g2}) contains  both  
positive and negative momentum $k$, 
the Fourier transform of the positive  $U$ frequency combination, 
$f_1$ in eq. (\ref{f1}), is:
\beq
\int_{-\infty}^{+\infty} f_1 e^{-ikz}dz =
{2i}\sinh\Bigl( {\pi\omega_\eta\over\alpha} \Bigr)
\Gamma\Bigl({i\omega_\eta\over\alpha}+1\Bigr)\tilde g_I(\eta),
\eeq
where $\kappa=\eta>0$, and $\Gamma$ is the Gamma function.
The Fourier transform of $f_2$ yields $\tilde g_{II}$.
Consequently, $\tilde g_I$ and $\tilde g_{II}$
correspond to positive 
frequency $U$-modes.
In the limit of $k_c\to \infty$, 
we conclude that 
outgoing radiation modes, 
$f_1$ and $f_2$, correspond in momentum space, to the modes
$\tilde g_I$ and $\tilde g_{II}$.

When dispersion in high momentum 
is present this correspondence is no longer  exact.
$g_I$ and $g_{II}$ will be modified near the horizon in
the regime $|z|< 1/k_c$.
Nevertheless, let us consider an outgoing wave packet,
and a time $t_0$ such that 
the  wave packet is localized 
in a domain sufficiently far away from the horizon.
At $t=t_0$ we can describe the outgoing Hawking photon
and its negative energy pair 
by the mode $f_1$, up to  a small correction.
The latter can be estimated by noting that in 
the momentum representation, large momentum amplitudes are  
suppressed like $1/k$. Thus, if we instead start at $t=t_0$ with 
the mode $G_\eta(k)\approx \tilde g_I$, with $\eta>0$,
it will correspond to $f_1$ up to
corrections of at most  $O(\lambda_c/M_{BH})$ 
where $\lambda_c = 2\pi/k_c$. 
Notice that unlike $f_1$, which is composed of both positive and 
negative parts, the corresponding mode $G_\eta(k)$ 
has a  positive
energy, i.e.,  an outgoing pair of thermal radiation
are described by $G_\eta(k)$ modes in
their ground state.  

To proceed with, let us now represent an outgoing Hawking photon
and its pair at 
$t=t_0$, by a wave packet of the 
exact modes  $G_{\eta}(k)$  with $\eta, \ k >0$:
\beq
A_g(k,t) = {1\over\sqrt{2\pi}}
\int_0^{\omega_c} e^{-(\omega-\omega_0)^2/\delta^2}
e^{-i\omega t} G_\eta (k) d\omega.
\eeq
$t_0$ is chosen such that $A_g(k,t_0)$ is mostly peaked at $k<k_c$, 
and $\omega_c$ is taken to be much larger then $\omega_0$.
We can then follow $A_g(k,t)$ as it evolves backwards in time, 
and observe the effects of dispersion. As
is shown in the following figures, the momentum gradually increases
and the wave packet, from which the radiation must have 
originated, is sharply peaked with $k\sim k_c$.

\medskip 
\begin{center}
\leavevmode
\epsfysize=8.0cm
\epsfbox{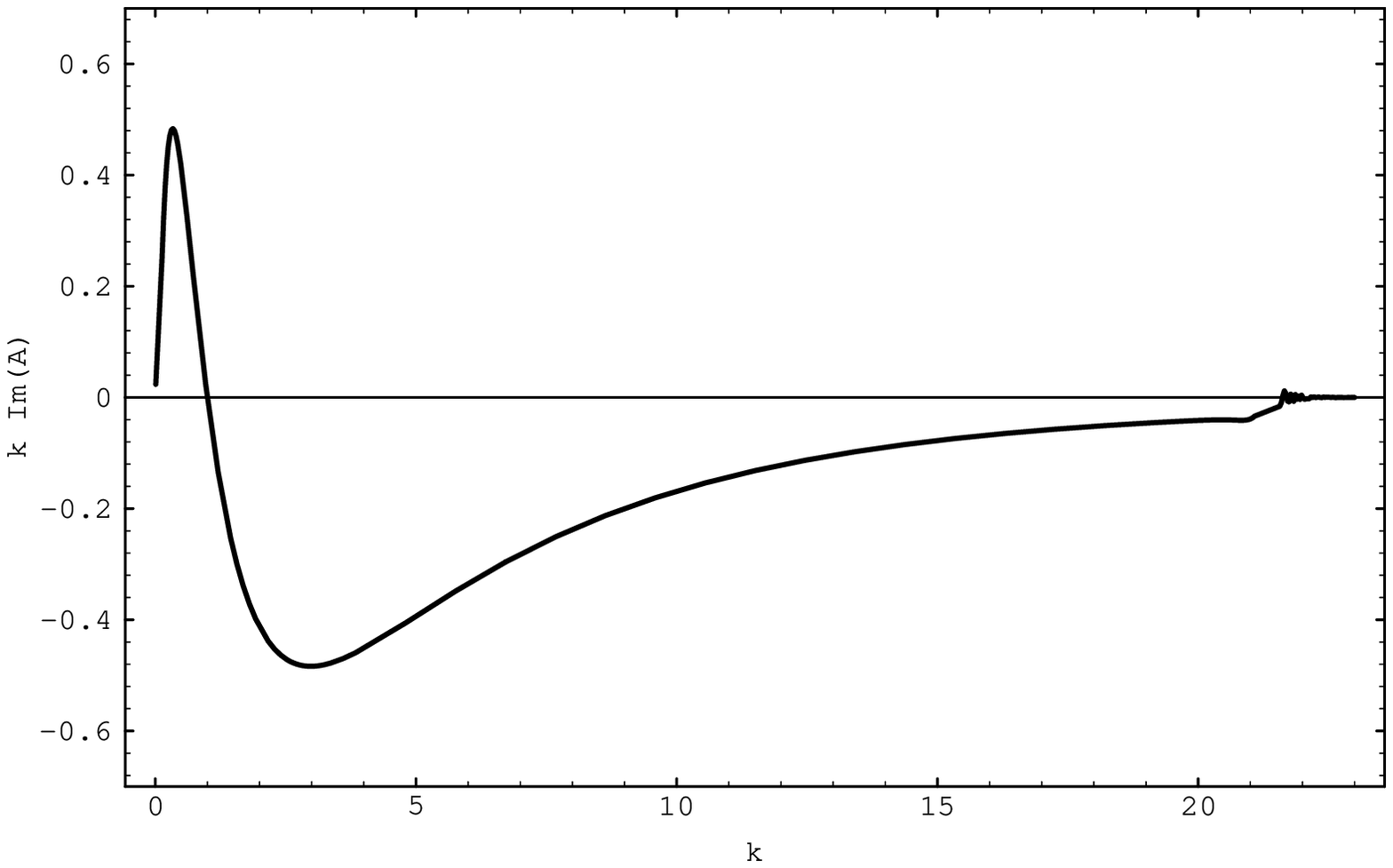}
\end{center}
\noindent
{\small Figure 1.  The imaginary part of $kA_g(k,t)$, at $t=0$,
 with $k_c =20,  \ \ \omega_0= \alpha = \delta =1, \  \ 
\Delta=0.5, \ \ \epsilon_c= k_c/\omega_0$} 
\medskip 

As can be seen from Figure 1, at this stage the dispersion 
causes only a minor modification at $k\approx k_c=20$, 
where the amplitude drops down to zero.

\medskip 
\begin{center}
\leavevmode
\epsfysize=8.0cm
\epsfbox{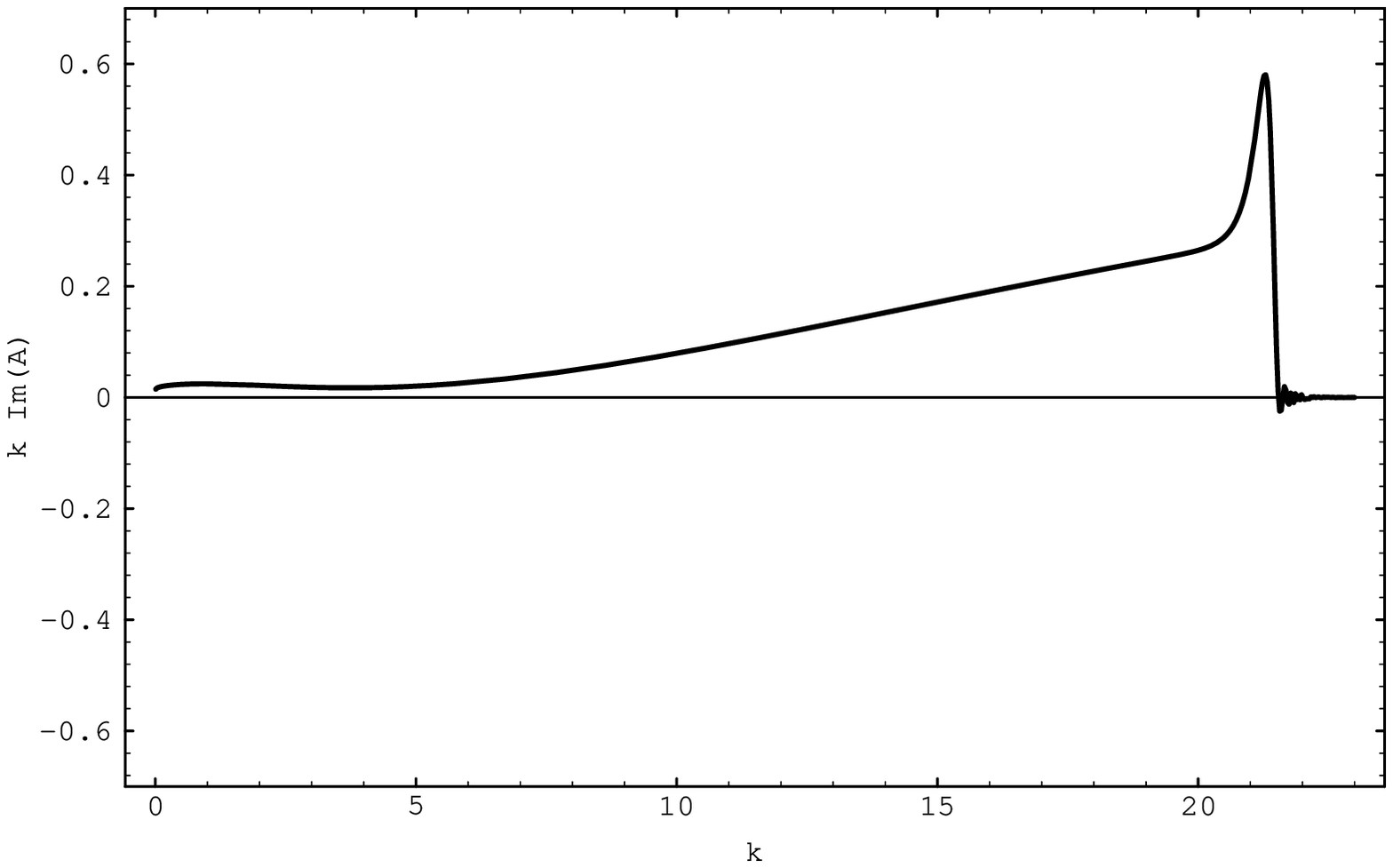}
\end{center}
\noindent
{\small Figure 2.   The wave packet at $t=-5$.}
\medskip 

\begin{center}
\leavevmode
\epsfysize=8.0cm
\epsfbox{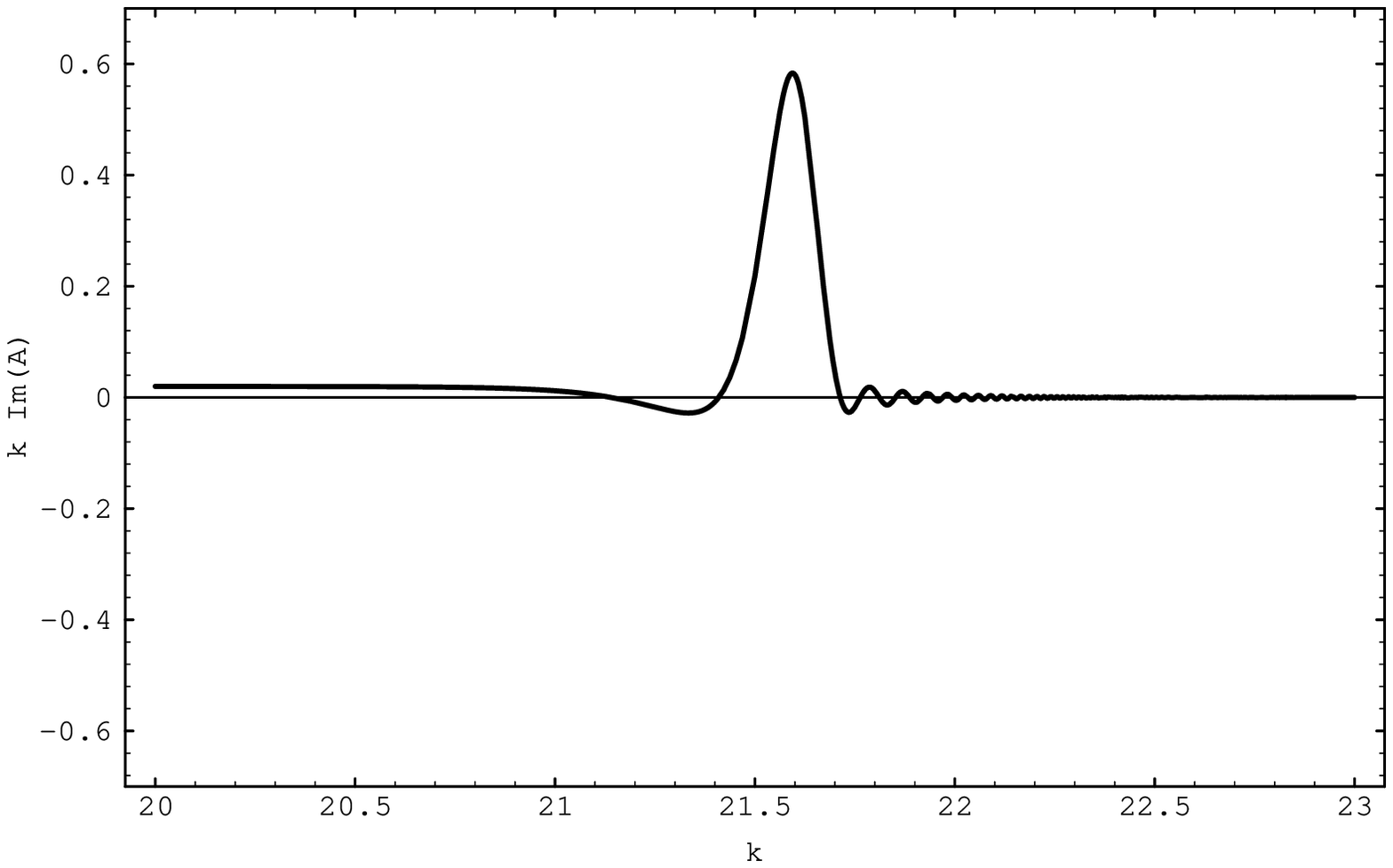}
\end{center}
\noindent
{\small Figure 3.   The wave packet  at $t=-10$.}
\medskip

In the last figure, the wave packet is completely   
localized within $\sim 3$ standard deviations, away from 
$k_c$.  It corresponds therefore to a wave packet of free
waves which interact very weakly  with the medium.
As we evolve $A_g$ even further back in time,
the wave  continues very slowly 
to shift towards larger $k$, and its width, $\Delta k$, 
narrows. 
For example at $t=-40$, the wave is localized at $k\approx 22$
with $\Delta k\approx 0.2$.
The corresponding spread in space is $\Delta z \sim 1/\Delta k$.
Therefore. the wave packet which gives rise to Hawking radiation, 
is at sufficiently early time 
much larger than the size of the black hole.

The value of $\epsilon_c$ in the simulations above has
been  chosen as $\epsilon_c = k_c/\omega_0$.
Since the transition occurs at  
$k\approx k_c$, and since for $k>k_c$ the frequency,
$\omega_0= k/\epsilon_c$, is  fixed  by the Hawking 
temperature, this choice corresponds to  
a minimal mismatch between the maximal dielectric constant,
$\epsilon_{max}\sim 1/\alpha z_{min} \sim k_c/\alpha$, 
for low momentum waves, and $\epsilon_c$
for  $k>k_c$.
This choice implies that the velocity of light (or  $g_{00}$)
drops down to a minimal value, $1/\epsilon_c$,
and remains constant for higher momenta. 
For other values, say $\epsilon_c=1$, free waves at 
$k\sim k_c$ have frequency $\omega\sim k_c$ which does
not correspond to the given Hawking frequency $\omega_0$.
In this case we find that the Hawking radiation originates from 
a wave packet in the same range of $k$ and the same amplitude
as above, but which is strongly oscillating as a function of $k$. 
Although only weakly interacting free 
modes with $\omega\sim k_c$ contribute, in the 
domain $k>k_c$, this
still effectively gives rise to the exact mode $G_\eta$ with 
low frequency $\omega=\omega_0$.
(A related  phenomena which acts in the reverse direction, 
yielding an effective high frequency mode from low 
frequency modes, was described in \cite{ultra}. 
We shall further describe this case elsewhere).

Before concluding let us comment on some open issues.
First we note that although the inclusion of spatial 
dispersion has  eliminated the singular electric polarization
at $z=0$,  we still can have $\veps\sim k_c/\alpha\sim 
\lambda_c /M_{BH}>>1$. 
In other words, the use of linear constitution relations is
questionable. 
Another question 
concerns the role of temporal dispersion, which has been so far 
ignored. In the present model, 
the latter does not seem to be essential in order to 
generate the Hawking radiation, since $\omega << k_c$ at any
time.  
Nevertheless, if we do assume a cutoff scale $\omega_c$, 
then the dielectric constant will typically be    
$\veps(t-t',z) \sim (z\alpha)^{-1}\sin (\omega_c (t-t'))/\omega_c$. 
Therefore, temporal dispersion induces a non-locality,
$t-t'\sim {1/\alpha z \omega_c}$, which greatly 
increases near the horizon. 
This phenomena may be relevant for understanding  
the final stages of black hole evaporation.

In conclusion, we have established that  outgoing radiation modes, 
originate in this model from nearly free 
modes, $G_\eta(k)$,  with $k\approx k_c$, which are weakly 
coupled to the medium. 
The latter are 
analogous  to  the Kruskal modes, 
$\exp(-i\tilde\omega U)$, in the standard picture, but 
do not contain wave numbers with $k\gg k_c$.
As a consequence, the initial state which gives rise
to Hawking radiation, most naturally turns out to be  
the ground state of nearly free  modes with $k\sim k_c$,
 in an effectively flat
background geometry.

\vskip 3 cm

{\bf Acknowledgment}
I would like to thank J. Anglin, S. Habib and P. W. Milonni
for valuable discussions.


\begin{thebibliography}{99}
\bibitem{dbh}
More general space-times can be easily obtained. 
For example the Schwarzschild geometry corresponds to the 
case $\veps_{zz}=\mu_{zz}=1, \ \ \veps_{xx}=\veps_{yy}=\mu_{xx}
=\mu_{yy} = r/r-r_0$, where the event horizon is at $r=r_0$.
\end{thebibliography}
\end{document}